\documentclass[11pt]{article}
\usepackage{vietnam,epsfig}
\def\3{\ss}

\def\Journal#1#2#3#4{{#1} {\bf #2}, #3 (#4)}

\def\PLB{{\em Phys. Lett.}  B}

\def\PRD{{\em Phys. Rev.} D}

\def\mco{\multicolumn}

\def\ra{\rightarrow}

\def\ko{K^0}

\def\be{\begin{equation}}
\def\ee{\end{equation}}
\def\bea{\begin{eqnarray}}
\def\eea{\end{eqnarray}}

\begin{document}
\vspace*{4cm}
\begin{center}
{\bf SUMMARY TALK: 

\smallskip
CHALLENGES IN PARTICLE ASTROPHYSICS}

\vspace{1cm}
HINRICH MEYER

\smallskip
{\sl Bergische Universit\"at Wuppertal, Fachbereich Physik, \\
Gaussstr. 20, D-42119 Wuppertal, Germany, \\ and 
DESY Hamburg, Notkestr. 85, 22607 Hamburg, Germany}

\end{center}

\section*{Introduction}
At this meeting we had 42 talks which covered most of the important
subjects of the field of particle astrophysics. In addition, on one
afternoon more details were presented in 5 parallel sessions with
another 42 shorter talks.

\bigskip
Particle Physics is very nicely described in terms of the rather
complete `Standard Model' with 6 quarks and 6 leptons interacting through
the exchange of photons, weak gauge bosons and gluons. So far no significant
deviations have surfaced. More and more details have been added to
complete the picture and there is no serious hint of a discrepancy
that would indicate new physics beyond the Standard Model. Only the
``Higgs'' is not found yet. But in the
last 10 years or so there have been great advances in the field of
Particle Astrophysics adding new information to particle physics while
simultaneously opening up new windows of observation in astrophysics
and astronomy.

\begin{enumerate}
\item[1.]
Neutrino oscillations have been discovered in studies of neutrinos
created in energy production in the sun and through cosmic ray interactions
in the earth atmosphere. Furthermore, neutrinos from SN~1987~A have been
detected, confirming qualitatively the basic expectations for a supernova
explosion. These important discoveries have been impressively
confirmed with oscillation observations of neutrinos from power reactors
and particle accelerators.
\item[2.]
Extensive studies performed with radiodetectors on satellites, on balloons
and on earth of the 2.7$^\circ$ K cosmic microwave background (CMB)
have revealed a flat universe ($\Omega = 1$) and supported by observations
of supernova of type 1A out to distances beyond $z = 1$, an energy
content of the universe dominated by Dark Energy and Dark Matter with
only a smallish fraction in the form of baryonic matter. There is also a
contribution of neutrinos with mass, although apparently on an even
much smaller level.
\item[3.]
TeV gamma ray experiments have detected by now more than 40 sources,
mostly of galactic origin but 10 are obviously from extragalactic sources.
\item[4.]
Searches for diffusive neutrinos from SN-explosions and for diffuse
infrared photons, both cosmological signals, have resulted in only
upper limits up to now but close to expected levels with prospects of
real observations in the next future.
\item[5.]
Searches for gravitational waves enter a new phase with the interferometers
LIGO, GEO600, VIRGO and TAMA taking regular data and pushing background levels
close to expectations.
\item[6.]
Searches for Dark Matter candidates approach sensitivity levels that
significantly constrain properties of new particles and of extreme
astrophysical objects.
\end{enumerate}

These success stories drive further developments of the field that
is based most importantly in very fundamental questions and problems
formulated long ago by great scientists in the somewhat distant past.
To name a few: Max Planck (1899), Victor Hess (1912), Albert Einstein (1916),
Wolfgang Pauli (1930), Fritz Zwicky (1933), Ettore Majorana (1937),
Peter Higgs (1962), Kenneth Greisen, Georgiy Zatsepin and Vadim Kuzmin (1966),
Andre Sakharov (1967), Bruno Pontecorvo (1968); of course a rather
incomplete list.

\medskip
Let me now try to cover some of the essentials of presentations at this
meeting, all on items at the forefront of particle astrophysics.

\section{Is there a Cutoff in the Allparticle CR -- Energy Spectrum?}
The most recent high statistics data come from the high resolution
fly's eye (HIRES) experiment. Overall there is within the limits of the
systematics rather good agreement with previous experiments. However, the
data show two features, firstly a shallow dip at $log_{10}(E)(eV) = 18.6$,
most likely due to $e^+e^-$ pair loss of protons and an apparent cutoff
at $log_{10}(E)(eV) = 19.6$ due to single pion photoproduction.
Both effects are consistently understood in terms of interactions of
primary protons of extragalactic origin with the $2.7^\circ K$ cosmic 
microwave background.

New experiments like Telescope Array  in Utah and most importantly
the AUGER experiment in Argentina are both much larger than previous
arrays. However, by combining the two techniques of scintillator
(Cherenkov) stations on the ground with the observation of the
fluorenscence light of the airshowers, one expects to improve not only
on statistics but in particular on the otherwise dominating systematic
uncertainties.

Based on the prediction by Askarian (1962) of detectable radio signals
from shower development in rock and ice, new experimental possibilities 
open up,
since in particular the acceptance area can be increased dramatically.
One aims for the detection of radio signals coming from high energy
neutrino interaction in antarctic ice (ANITA) or even in moon rock
(NuMoon). The aim is of course GZK neutrinos that would consist
of a beautiful new cosmological signal due to cosmic ray protons
interacting with 2.7$^\circ K$ photons everywhere in the distant universe,
as expected from the most straighforward interpretation of the HIRES result.

\section{Dark Matter}
One obvious interpretation of rotation curves of spiral galaxies is the
existence of astronomical objects of subsolar mass which can be detected
through their action as gravitational lenses on background stars.
Two larger collaborations, MACHO and EROS, observed millions of stars
in the large (small) `Magellanic Clouds' to search for lensing events
due to dark objects. The result of this monumental effort is now quite
convincing; so called MACHOS in the mass range ($10^{-7} - $ a few) solar
masses do not provide for the dark matter of our galaxy, all observed
lensing events are of known origin, new dark astronomical objects are
not required. Too bad.

New extended observations of dwarf spheroidal galaxies in the
more local environment of our galaxy, some of them only recently discovered,
in large ongoing surveys, e.g. SDSS. Large telescopes of the 8 -- 10 meter 
class,
equipped with  multi object spectographs, have allowed for detailed studies
of the dynamical behaviour of stars in dwarf ellipticals. The results
are very puzzling, the galaxies seem to be dominated by dark matter, but all
with about $4 \times 10^7$ solar masses in DM, despite quite a range in
mass/light. In standard models, where large galaxies result from
merging many smaller ones over cosmic time one would expect to see at least 
in some of them
 (strong) tidal effects, but this is not what is observed. Also alternative
theories like e.g. MOND seem to fail. On the other hand MOND is very 
successful in describing rotation curves of spiral galaxies, small and
large, in even fine detail. And MOND has somehow become more acceptable
since a relativistically invariant version, TeVeS, with classical MOND as
the nonrelativistic limit has been found by Bekenstein (2003). The
dynamical behaviour of galaxies in clusters, however, cannot be 
quantitatively described by MOND, massive neutrinos of cosmological
origin with mass of about 2 eV (at least one species) might cure the
defect.

Will laboratory efforts be more successful? Over the last 20 years there
have  been great advances in the suppression of background, by roughly
a factor of $10^6$. But also new ingenious detection methods have been
developed. No convincing dark matter candidates (Whimp) have been found
so far. So the trend is for larger detector masses without compromising the 
background level. The aim is to come below scattering cross section
levels for mass ranges indicated by e.g. supersymmetric extensions
of the standard model. Will classical particle physics, in fact LHC,
provide candidates? Even if that's the case they still need to
be observed in DM-search experiments outside accelerators to solve the
DM-enigma of astrophysics.

\section{Low Energy Neutrinos}
Neutrinos already show a rather complicated phenomenology and very much is
still to be discovered. Of foremost importance is a measurement of
theta (1,3) which is certainly much smaller than both theta 
(2,3, 1,2). To get a value for the (1,3) mixing angle large efforts are
required and several experiments are underway. Double Chooze may be next with a
result but if they get a limit only (pessimistically), at least the large
scale multinational effort at Daya Bay is needed. A small value of theta
(1,3), however, will make it very hard to find CP violation effects.
It sounds like a long way still.
But it must as well be admitted that any reasonable understanding of the
mass spectrum of (elementary) particles seems far below the horizon.
In particular: why are neutrino masses so small?
Indeed, at this meeting we heard the phrase ``there is
a problem of mass'', quite strongly expressed by several speakers.

\section{Early Universe}
At least back to MeV temperatures where nucleosynthesis happened
we think we have a pretty complete and consistent picture of the
development of the universe. But there are still some irritating results.
It seems rather difficult to fix the $^4He$ content in the universe
with sufficient precision, $^2H$ measurements are only very few
and in particular the level of $^7Li$ did not fit by about a factor
of 2. It is observed in old metal deficient stars and recently it has
been claimed that diffusion plus turbulent mixture may account for
about this factor of 2 reduction of $^7Li$. But then a new problem
seems to appear, $^6Li$ in old stars may be much too abundant to be
of primordial origin.

While CMB observations are a beautiful quantitative tool for cosmology by now
and with even better observations to come (Planck satellite) one may pose
the question  -- is 
the light element production casting doubts on the
correct understanding of the early universe?

\section{TeV Gamma Astronomy}
In the `90ties both Whipple and HEGRA made ground breaking discoveries
in the particle astrophysics field. The overwhelming drive comes from the
attempt to solve the question of the origin of cosmic rays.
High energy TeV sources were discovered, galactic and extragalactic.
Two new experiments, H.E.S.S. in Namibia and MAGIC on La Palma
are follow up experiments of
HEGRA. Both are very successful and lead the field by now.
Larger mirrors and large acceptance cameras led to the discovery of 
many new sources, mostly located inside our galaxy. The source number
count is now more
than 40. At least 5 different classes have been established, SN remnants,
binary star systems, extended emission regions and pulsars.
The extragalactic sources are all Blazars, a special subclass of active
galactic nuclei (AGN). Further M87, long suspected to be an 
important extragalactic contributor to cosmic rays, has by now been firmly
established. Photon energy spectra extend up to nearly 100 TeV. Energy spectra
from Blazars have allowed for a strong constraint on the intergalactic 
(infrared)
photon field to a level just compatible with the results from galaxy
counts. No extra sources seem to be required and earlier larger contributions
to the infrared photon flux seem now rather safely excluded.

\section{Planck Mass}
In discussions of attempts to get a theory of quantum gravitation the Planck 
Mass $M_{P\ell}$  sets the scale in a number of relations. It is
appropriate I think, to go back and have a look at the original
publication of Planck which appeared in the `Sitzungsberichte der
preu\3ischen Akademie der Wissenschaften' in 1899. Planck gave a series
of five reports ``\"uber irreversible Strahlungsvorg\"ange'' (on irreversible
radiation events). \S 26, the last paragraph of the whole series, has
the title ``Nat\"urliche Masseeinheiten'' (natural units). Here he
introduces fundamental units for length, time and mass (and temperature)
as combinations of the gravitational constant ($f$), the velocity
of light ($c$) and a constant ($b$) which today is Planck's $h$,
that he introduced 
in the foregoing reports. His considerations are truly astonishing as
he argues that these units are independent of specific bodies or substances
and keep their meaning for all times and for all, even extraterrestrial and
nonhuman cultures, as long as the laws of gravitation, light propagation
in vacuum and both laws (Haupts\"atze) of the theory of heat remain valid.
Determined by even very different intelligent beings they have to have the same
values.
Did he read Jules Verne?

\medskip
Finally, I cannot resist to comment on the name of a new all
purpose underground laboratory discussed by B. Sadoulet, ``DUSEL'', actually
a proper German word whith the meaning fluke (somehow undeserved).

\section*{Acknowledgements:}
I would like to thank Jacques Dumarchez for the kind invitation to this
meeting. Exellent support in the preparation of the summary was
provided by Gabriel Chardin and Florian Goebel. Sincere thanks go to
Jean Tran Thanh Van for his tireless effort to create Rencontres du Vietnam
and his kind friendship over about 40 years.

\end{document}